\documentclass[aps,prl,twocolumn,superscriptaddress,showpacs]{revtex4}
\usepackage{hyperref}
\usepackage[dvips]{graphicx}
\usepackage{amssymb,amsmath,amsfonts}

\begin{document}

%\preprint{}

\title{Comment on "Tunable Band Gaps in Bilayer Graphene-BN Heterostructures"}

%%%%%%%%%%%%%%%%%%%%%%%%%%%%%%%%%%%%%%%%%%%%%%%%%%%%%%%%%%%%%%%%%%%%%
%% The document title should be given as usual. Some journals require
%% a running title from the author: this should be supplied as an
%% optional argument to \title.
%%%%%%%%%%%%%%%%%%%%%%%%%%%%%%%%%%%%%%%%%%%%%%%%%%%%%%%%%%%%%%%%%%%%%

\author{J. S\l awi\'{n}ska}
\affiliation{Theoretical Physics Department II, University of Lodz, Pomorska 149/153, 90-236 Lodz, Poland}
\author{I. Zasada}
\affiliation{Solid State Physics Department,University of Lodz, Pomorska 149/153, 90-236 Lodz, Poland}
\author{P. Kosi\'{n}ski}
\affiliation{Theoretical Physics Department II, University of Lodz, Pomorska 149/153, 90-236 Lodz, Poland}
\author{Z. Klusek}
\affiliation{Solid State Physics Department,University of Lodz, Pomorska 149/153, 90-236 Lodz, Poland}

%\thanks{}

\vspace{.15in}

\begin{abstract}
We study the electronic properties of h-BN/graphene/h-BN ABC-stacked trilayer systems using tight binding and DFT methods. We comment on the recent work of Ramasubramaniam \textit{et al.} (arxiv:1011.2489) whose results seem to be in disagreement with our recent calculations. Detailed analysis reaffirms our previous conclusions.

\end{abstract}

\maketitle


%\section{Introduction}
Following our recent paper \cite{nasza2}, Ramasubramaniam \textit{et al.} have investigated the electronic properties of h-BN/graphene/h-BN sandwiched structure and generalized it on the configurations containing bilayer graphene (BLG) \cite{ramasubramaniam}. They found, however, that their DFT calculations are not consistent with our results \cite{nasza2} obtained  by means of combined tight binding (TB) and ab initio methods. While overall valuable, we believe that this article does not invalidate our model and conclusions. The purpose of this paper is to explain the contradictions and to clarify different interpretations.

The previous analysis of the electronic properties of graphene/h-BN structures \cite{nasza1, nasza2} was based on the minimal tight binding model with parameters determined by density functional theory (DFT) calculations. As far as we know, the basis containing localized orbitals is quite reasonable in calculations for materials consisting of boron, carbon and nitrogen. Tight binding method allows to obtain correct spectra of graphene and h-BN sheets in the whole Brillouin zone when three nearest neighbors are included in the model, while the qualitative description of  dispersion relations in the vicinity of K/K' points is possible even when only nearest neighbor model is considered. We found the small discrepancies in the TB curve fitted to DFT results \cite{nasza2}, but this fact cannot significantly influence the properties of the system, especially because the Dirac cones are preserved in DFT band structure.

It has been assumed in the TB model that the positions of atoms within each layer are fixed: the same geometry was used in DFT calculations, i.e. no relaxation was allowed. For completeness, we have calculated the spectra of relaxed structure using exactly the same parameters as in the work of Ramasubramaniam (see also Methods in Ref. \cite{ramasubramaniam}). It is clear that the small differences in lattice structures do not strongly affect the $\pi$ bands, since the Dirac cones are still preserved in the ABC trilayer containing N-C dimers.
In addition, we would  like to stress that DFT calculations described in Refs \cite{nasza1, nasza2} were performed under almost the same assumptions as those for graphene on h-BN substrate \cite{hbn} and the results given in Ref.\cite{hbn} have been well reproduced. Therefore it can be expected that the same parameters will give correct results for bilayer and trilayer graphene/h-BN configurations.

The next important issue is to explain the statement in the footnote 26 in Ref.\cite{ramasubramaniam}. Figure 3 in Ref. \cite{nasza2} shows the variation of the band gap with the effective value of energy difference between the layers. The effective value means exactly the U parameter in the TB hamiltonian (equation (1) in Ref. \cite{nasza2}) and it is not equivalent to the field in the inputs of DFT calculations, since the screening effects are not included in TB model. The value of U related to each DFT band structure  in Refs \cite{nasza1, nasza2} is estimated by analysis of band shifts which are predicted by TB model: the shift of bands in the spectra determines the exact value of parameter U used in TB hamiltonian. This means that the value 5 V/nm in Ref. \cite{ramasubramaniam} is equivalent to about 1.15 eV in the scale of Fig.3 of Ref.\cite{nasza2}. It can be easily found that for this value of U the predicted band gap is about 40 meV , which is in agreement with the results of Ramasubramaniam shown in Fig.5 (e) of Ref. \cite{ramasubramaniam}. 

Moreover, it is explicitly stated in the last section of Ref. \cite{nasza2} (illustrated in Fig. 3 therein, where the system is compared to the freestanding BLG), that, in practice, the opening of theoretically predicted band gap of about 230 meV may be very difficult to achieve, because, for this system, the very large or even unphysical electric fields are needed. It is also worthwhile to note that similar highly symmetric ABC configuration containing B-C dimers was rejected as less promising, because the band gap can be at most 50 meV and cannot be larger even theoretically. Therefore, the band gaps in both B-C and N-C ABC trilayers hardly depend on the applied electric field. On the other hand, the most important feature of the considered trilayers is the presence of the Dirac cones in the vicinity of the Briluoin zone corners and the fact that the band gap can be created.
 
\begin{figure}
\includegraphics[width=0.9\columnwidth]{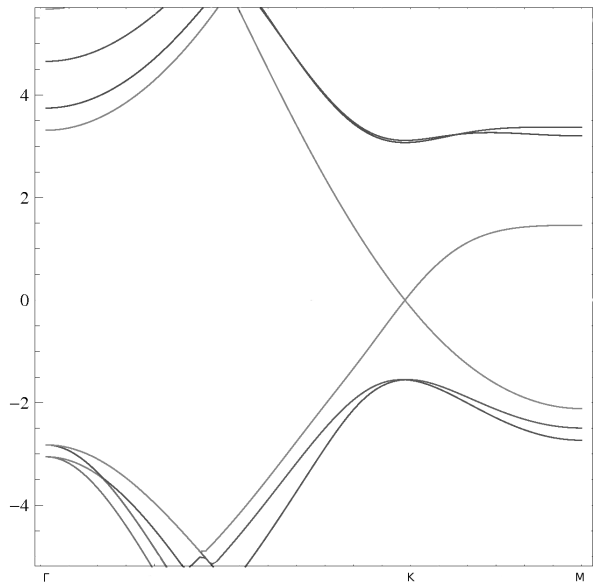}
\caption{Band structure of the relaxed ABC h-BN/graphene/h-BN trilayer containing N-C dimers.}
\end{figure}

Finally, contrary to the statement of authors of Ref.\cite{ramasubramaniam}, we observe that, in the band structures shown in Figs 5 (a)-(c) (ABA configurations) and Fig. 5 (f) (ABC with different dimers) in Ref.\cite{ramasubramaniam}, the linear dispersion is not preserved near K/K' points due to the symmetry breaking between graphene sublattices. It is typical of the graphene/h-BN systems that the mass of Dirac fermions increases with the value of a band gap \cite{semenoff}; thus in all cases where the band gap is present the $\pi$ bands should be parabolic. The effective mass is finite, which is a fundamental difference when compared to the freestanding monolayer graphene. We have performed the calculations for the case of relaxed structures shown in Figs 5 (d)-(e) of Ref. \cite{ramasubramaniam} and we found no gap between valence and conduction bands (Fig. 1). Therefore, the statement that the band gaps are in range of 10-100 meV is not clear to us.

It should be stressed that, indeed, the bilayer graphene inserted between the h-BN sheets can provide a system of immediate practical relevance for graphene - based devices, especially in view of recent experiments \cite{substrat_hbn}. The mechanism of band gap tuning in h-BN/BLG/h-BN sandwich is very similar to that in the freestanding BLG and the sensitivity to the stacking is not as important as in the configurations containing monolayer graphene (graphene is randomly stacked on bulk h-BN \cite{substrat_hbn}). On the other hand, the band structure manipulations in the ABC systems with single layer graphene are unique and could offer new possibilities of unusual applications. Unfortunately, the exact ABC stacking is needed for the full realization of band gap tuning in this system. The lattice constants should be precisely matched, which is not certain until now \cite{usachov}.

\begin{acknowledgements}
This work is financially supported by Polish Ministry of Science and Higher Education in the frame of Grant No. N~N202~204737. One of us (J.S.) acknowledges support from the European Social Fund implemented under the Human Capital Operational Programme (POKL), Project: D-RIM. 
\end{acknowledgements}
\end{document}